# Biomimetic Engineering of a Fortified Ice Composite with Enhanced Mechanical Properties

Chen ADAR[1#], Yulia BARON[1#], Baruch ROFMAN[2#], Maya BAR-DOLEV[1,3], Liat BAHARI[1], Victor YASHUNSKY[1,4], Vera SIROTINSKAYA[1], Oded SHOSEYOV[5], Ido BRASLAVSKY[1]*

1 Institute of Biochemistry, Food Science, and Nutrition, Robert H. Smith Faculty of Agriculture, Food and Environment, The Hebrew University of Jerusalem, Rehovot 7610001, Israel

2 Swiss Plasma Center, Ecole Polytechnique Fédérale de Lausanne, Lausanne CH-1015, Switzerland

3 Faculty of Biotechnology and Food Engineering, Technion, Haifa 3200003, Israel

4 The Swiss Institute for Dryland Environmental and Energy Research, The Jacob Blaustein Institutes for Desert Research, Ben-Gurion University of the Negev, Sede Boqer Campus, 84990 Midreshet Ben-Gurion, Israel

5 Department of Plant Sciences and Genetics in Agriculture, Robert H. Smith Faculty of Agriculture, Food and Environment, The Hebrew University of Jerusalem, Rehovot 7610001, Israel

[#]Equal contribution, *corresponding author

**KEYWORDS**

ice-binding proteins; antifreeze proteins; ice affinity purification; cellulose nanocrystals; Carbohydrate-Binding Module; ice templating; directional freezing.

**Abstract**

This work presents BioPykrete, a new sustainable bio-composite material created from ice, nano-crystalline cellulose (CNC), and a tailor-made chimera protein designed to bind the two together. We developed and produced the chimera protein by linking AFPIII, an ice-binding protein, with CBM3a, a CNC-binding protein. As the suspension freezes, the CNC chains self-organize into a reinforcing network between the ice crystals. This structural enhancement limits crack propagation to typical pore sizes, allowing BioPykrete to avoid the brittle and sudden failure commonly associated with ice. Instead, it exhibits an elastic-like response to stress, making it suitable for construction and engineering applications. With compressive strength comparable with concrete, BioPykrete offers a sustainable and biodegradable alternative to construction materials suitable for the harsh arctic regions of the world where traditional methods are ineffective, and resources are scarce. Engineering chimera proteins with specific affinity to more than a single material type may help improve or tailor the properties of other composite materials.

## 1. Introduction

In the arctic regions of the world, viable construction materials are scarce and expensive commodities, while the construction itself is challenging and requires significant care, resources, and energy. Concrete, the most common go-to solution, must be kept above five °C during the casting and curing stage to reach its full strength. This entails the erection of insulating structures protecting the concrete, real-time monitoring of temperature and humidity in the mix, and the use of accelerating additives. These are not magic bullets; they increase a project's complexity, cost, time, and risk. Unless specialized mitigation is applied, the concrete would lose up to 50% of its mechanical integrity after 30 thermal cycles between -40 °C and 0 °C,[1–3] leading to a higher need for infrastructure maintenance and reconstruction. Thus, due to the harsh conditions and complications, Arctic construction projects are more costly, resulting in a bigger carbon footprint than average.

Ice offers a readily available, low-cost, sustainable alternative for arctic construction. However, although ice only increases in strength at lower temperatures,[4] it is also brittle. It might experience a sudden and catastrophic failure, severely limiting its usability for construction applications.[5,6]

Brittle materials, such as ice, are often rigid and strong but are especially sensitive to the combination of small notches and tensile stresses, resulting in rapid crack propagation followed by catastrophic structural failure.[7] Therefore, such materials are rarely used for structural applications. However, introducing reinforcing elements that impede the propagation of fractures is known to change the failure mechanism to a non-linear deformation similar to that of ductile materials.[8]

Organic-based reinforcement can be used to turn ice into a sustainable structural material. Cellulose-based additives are excellent candidates as they offer low-cost, non-toxic, biodegradable solutions from renewable resources. "Pykrete", a wood pulp and ice composite developed in the '40s [9], is an example of a cellulose-based ice composite. The term "Pykrete" has also been used for other cellulose-ice composites like sawdust, all showing improved mechanical properties relative to pure ice.[10–13] Despite improvements in Pykrete composition and processing, its real-life application remains limited [10,14]. Pykrete's mechanical properties depend on the fibers' size and density [11]. The cellulose fibers used in Pykrete are on the millimeter to micron scale and contain large amorphous regions with variable shapes. This cellulose structure leads to the fibers having low mechanical strength and limited self-assembly and re-orientation during freezing, which influences their integration into the brittle matrix.[15]

Cellulose is organized in hierarchical structures which are either amorphous or crystalline. Specifically, cellulose nanocrystals (CNC) are stiff, rod-like nanoparticles of cellulose segments that are extractable from various biological sources, usually by mechanical and chemical treatments.[16] While bulk cellulose has significant disordered fractions, CNC constitutes the crystalline fraction of cellulose (50% - 80%) and exhibits a high aspect ratio (1:100), large specific surface area, high tensile strength, and elastic modulus, along with reactive surfaces. These properties make CNC a desirable additive for the reinforcement of composites. Indeed, implementing CNC into polymers resulted in composites with superior mechanical properties (reviewed in [16–20]). With these properties, CNC may serve as an excellent reinforcement matrix for ice.

To further improve the properties of the ice-cellulose composite, we introduce a method to adhere them together at the molecular level. We present a chimera protein engineered by combining a cellulose-binding protein (CBM) and an ice-binding protein (IBP). This chimera is a biological coupling agent, or "adhesive" that binds the CNC directly to the ice. We produced the chimera by fusing two genes coding for independent and unrelated natural proteins. The first protein is an antifreeze protein (AFP), a type of IBP that adheres irreversibly to ice crystals.[21]

The second protein is a carbohydrate-binding module type 3a (CBM3a) of the cellulosome-integrating protein CipA that adheres specifically to cellulose.[22]

We designed and produced the fusion protein CMB3a-AFPIII, first on a small scale and then on a large scale. We tested its activity and stability, ensuring it can bind to CNC and ice independently. Next, we fabricated three new CNC-ice composites with increasing complexity and investigated their morphology, mechanical strength, and engineering viability. Our work demonstrates that by combining the CNC network with the chimera protein to create BioPykrete, we gain a 10-times increase in compressional strength, a 70-times increase in energy to failure, and ductile-like failure mechanics.

## 2. Experimental Methods and Materials

### 2.1 CNC Suspension

The CNC suspension (at pH 5.6) was a gift from Melodea (Rehovot, Israel). Enzymes for cloning were obtained from New England Biolabs (MA, USA). Unless mentioned otherwise, all other reagents were purchased from Sigma Aldrich (Merck Group).

### 2.2 Engineering of CBM3a-AFPIII chimera

A detailed description of the production of the CBM3a-AFPIII genes and the gene and protein sequences are available in the supporting information. Briefly, the pET28 plasmid containing the gene coding for CBM3a CipA from the Cellulosomal scaffolding of *Clostridium* Thermo*cellum* [22] was generously provided by Ed Bayer (Weizmann Institute of Science, Rehovot, Israel). A plasmid containing the gene coding for AFPIII, the rQAE m1.1 isoform from ocean pout [23] [24] was obtained from PL Davies (Queen's University, Ontario, Canada). The AFPIII gene was amplified by polymerase chain reaction (PCR) to incorporate either a short linker of two amino acids or a long linker of 10 amino acids upstream to the AFPIII sequence. The PCR product was digested and cloned into the pET28 plasmid upstream to the CBM3a gene. The cloned plasmids were transformed to BL21(DE3)PlysS *E. coli* (Novagen) for protein synthesis.

**2.3   CBM3a-AFPIII production**

**2.3.1   Small-scale protein production**

The recombinant protein, with either the short or the long linker, was expressed in *Escherichia coli* BL21 cells in Luria-Bertani (LB) broth supplemented with 100 μg L$^{-1}$ of kanamycin. The growth temperature was kept at 37 °C until the optical density of the culture at 595 nm (OD$_{595}$), indicative of cell concentration, reached 0.6. The temperature was lowered to 16 °C, and the culture was supplemented with one mM of isopropyl β-D-1-thiogalactopyranoside (IPTG) to include protein production. The bacterial culture was allowed to grow overnight at 16 °C. Recombinant *E. Coli* cells were harvested by centrifugation at 6000 rpm and at 4 °C for 15 min. The cells were pelleted and resuspended in 40 ml Binding Buffer (5 mM imidazole, 50 mM Tris-HCl, 150 mM NaCl, pH 7.6). The suspension was kept at -80 °C until use.

**2.3.2   Large-scale protein production**

Recombinant CBM3a-AFPIII protein was expressed in a stirred tank fermenter of 1 L or 5 L (Applikon Biotechnology, The Netherlands) for a fed-batch fermentation process. The fermentation procedure's development was described in detail by Sirotinskaya *et al.* [25]. The bacteria were grown at 37 °C under controlled physical and chemical parameters. When the culture reached an OD$_{595}$ of 0.6, the temperature was lowered to 15-20 °C, and one mM IPTG was added. The typical fermentation duration was 20-24 hours, and final OD$_{595}$ values reached 60 to 80. Cells were harvested by centrifugation at 4 °C at 4500 rpm for 45 min and stored at –80 °C until purification. The control proteins, AFPIII and CBM3a, were prepared following the same procedures except for the cultivation temperature, which was 37 °C throughout the expression, and the cultivation time, which was 4-6 hours. A more detailed description of the production of the chimera protein is available in the SI.

### 2.3.3 Protein purification

The proteins were purified by NI-NTA affinity chromatography as described in Sirotinskaya *et al.* [25]. SDS-PAGE evaluated protein purification yields and molecular weights. Proteins were concentrated using Vivaspin ultrafiltration columns (Sartorius), and the final protein concentrations were determined using the micro-BCA Protein Assay (Thermo Scientific, cat#23235).

### 2.4    Characterization of CMB3a-AFPIII

### 2.4.1    Thermal hysteresis assay

The thermal hysteresis (TH) measurements were conducted using a custom computer-controlled nanoliter osmometer. [26] The procedure is described in detail in [27]. A few nL drops of protein solution were injected into an oil-filled cavity in a metal sample holder. The holder was then placed on a cooling stage and rapidly cooled until nucleation occurred, typically between -30 °C and -35 °C. The frozen drop was slowly warmed until a single ice crystal remained. The crystal was carefully allowed to melt to a diameter of 10-20 µm, and its melting temperature was recorded. Then, the temperature was slightly lowered (to stabilize the crystal and prevent further melting), and the crystal was incubated for 10 minutes to allow protein to accumulate on its surfaces. After the incubation period, the crystal was cooled at a constant rate of 0.0025 °C s$^{-1}$ until ice growth (typically in a "burst" like manner) was observed. The difference between the melting and the burst-freezing temperatures is the TH of the sample. At least three independent TH measurements were taken for each protein concentration, as presented in Figure 2.

### 2.4.2 Ice affinity

Ice affinity experiments were performed following, with slight modifications, the methodology developed by Kuiper et al.[28]. After Ni-NTA purification, 5-10 mg of protein were diluted in 1 L of 50 mM Tris-HCl buffer at pH 7.6 and cooled to 4 °C. A hollow, cold finger was cooled using a circulation chiller to approximately -5 °C and submerged in double-distilled water (DDW) to facilitate rapid ice formation on its surface. Once the ice layer reached a thickness of around 1 mm, the cold finger was removed from the water and placed into the cooled protein solution. The temperature of the cold finger was set to -4 °C while the solution was continuously stirred using a magnetic stirrer. After 10 minutes of temperature equilibration, the temperature of the cold finger increased from -4 °C to -7 °C at a rate of 0.008 °C min$^{-1}$. Subsequently, the ice on the cold finger was washed with DDW to remove any nonspecifically attached proteins to its outer surface. The ice was then allowed to melt at room temperature, and the volume of the melted ice was measured. A concentrated Tris-HCl buffer was added to adjust the solution to a final concentration of 50 mM, assuming that the salt was excluded entirely during ice growth and that only purified water retained the protein. The ice fraction was concentrated using a centrifugal concentrator (Vivaspin 20, 10000 MWCO, Sartorius). Samples from the melted ice fraction and the remaining solution were analyzed using SDS–PAGE.

### 2.4.3 CNC binding assay

The binding capability of CBM3a-AFPIII to CNC was evaluated following a previously described procedure [29] with some modifications. Thirty µL of either purified CBM3a-AFPIII or only AFPIII solution as a control, at a concentration of 50 µM, were mixed with 30 µg of CNC (Avicel PH-101) in 50 mM Tris-HCl, 150 mM NaCl, pH 7.6 (TBS) at a final volume of 200 µL. After one hour of incubation at 4 °C, to separate the unbound protein left in the solution from the bound protein in the pellet, the samples were centrifuged at 14000 rpm for two min while still maintained at 4 °C. The pellets were washed twice with 200 µl TBS supplemented with Tween 20 to prevent non-specific interactions. After each wash, the supernatant was separated by centrifugation for two min at 4 °C and 14000 rpm and collected. The washed pellets were suspended in 45 µl TBS. Samples from the supernatants, suspended pellets, and the wash stages were supplemented with a sample buffer containing 10% SDS and boiled for 10 min. This procedure unfolded the proteins, releasing them from the cellulose into the solution.

### 2.5 Preparation of CNC: CBM3a-AFP III: ICE composites

A 3% W/V CNC suspension at pH 5.6 was centrifuged at 4500 rpm for 5 min at room temperature, sonicated, and adjusted to pH 7.5 in Tris-HCl buffer. The final suspension contained 2% CNC and 17 mM Tris at pH 7.5. Either CBM3a-AFPIII or control proteins were added to the CNC suspensions to a final concentration of 50 µM. In the pure composite, DDW was added in equivalent volume.

We used custom cups with copper bottoms and insulating Perspex walls for improved unidirectional heat transfer (Figure S2-a). The cups were loaded with either 10 ml suspension for the compression strength tests or 3 ml suspension for the SEM analysis and placed on a cooling stage. The cooling rate was regulated using a LabVIEW-based PID feedback loop, which controlled the flow rate of liquid nitrogen through a cryogenic solenoid valve (VCW31-5C-5-02N-C-Q, SMS, Japan). The solenoid valve was connected to a liquid nitrogen pressure tank, ensuring a steady coolant supply. Temperature monitoring of the stage was performed using a thermistor mounted on the cold stage. The PID controller used the thermistor readings to dynamically adjust the valve operation, allowing precise control of the cooling rate.[30]

To allow for directional freezing at a constant rate, the temperature of the cooling stage was lowered at a rate of 5 °C min$^{-1}$ (for the mechanical strength tests) or 10 °C min$^{-1}$ (for SEM imaging) until the entire suspension was frozen. The top of the suspension froze when the bottom of the sample reached typically − 80 °C. The ice composites for the strength analysis were stored at − 20 °C until use, while the composites grown for SEM imaging were lyophilized by connecting the growth chamber directly to a vacuum pump.

### 2.6 Scanning electron microscopy

The morphology of the CNC-ice composites was assessed using SEM (JEOL, JSM-7800F equipped with a Quorum PP3010T system). The lyophilized samples were cross-linked to improve their stability for SEM. The cross-linking was conducted by Melodea Ltd Company and included soaking in ethanol, maleic anhydride, glycerol, and castor oil, followed by overnight curing at 120 °C. As preparation for the SEM analysis, the foams were sliced perpendicular to the ice growth direction, mounted on a carbon tape, and coated with a 1 nm layer of Iridium (Quorum Q150T ES). The microphotographs were recorded using the high-resolution scanning electron microscope (JEOL, JSM-7800F). The images were taken with an accelerating voltage of 1-5 kV.

## 2.7 Mechanical testing setup

We prepared samples from four types of composites: pure ice made of DDW, DDW with 2% CNC, DDW with 2% CNC and 50 µM AFPIII, and DDW with 2% CNC and 50 µM AFPIII-CBM3a chimera protein. We produced cylindrical 1 x 5 cm (height x diameter) in size that were directionally frozen in the dedicated setup.

We measured the compression strength of the frozen samples with an adopted Instron device (3345, USA). The samples were placed between two metal plates and compressed along the ice growth direction until failure was reached. It was essential to keep the samples frozen and their temperature nearly constant during the compression experiment. Therefore, the entire setup was enclosed in a transparent plastic box, and the two metal plates of the Instron were wrapped in an insulating material and cooled by dry ice. Figures S2.c-d show the modifications we made to the device. Figures S2.e(I-III) show a typical compression test. Initially, the sample is slightly loaded until most of the surface area comes into contact with compression plates on both sides. Then, it is compressed at a constant rate until it reaches mechanical failure, while the strain-stress response is recorded. Depending on their strength, the samples were tested using either a 100 N or a 5000 N load cell. The compression was stopped automatically when the sample broke or manually after the sample's compressive stress reached its maximum. Figure S2.e.II shows a typical compressional failure in which the bottom and top spread outwards radially, resulting in localized crack formation.

## 3. Results and discussion

### 3.1 Design and production of CBM3a-AFPIII protein chimera

We selected the protein subunits for cloning the recombinant chimeric protein: (1) Only IBPs and CBMs with well-characterized structures (*i.e.,* solved 3D structure) and activities were considered. (2) Proteins with post-translational modifications were not considered to allow production in bacteria. (3) Small and soluble proteins were preferred for high-yield expression in bacteria and production scalability. (4) Proteins with many disulfide bonds were excluded to reduce potential oligomerization and aggregation. (5) Proteins with "floppy" or intrinsically unfolded sections were not considered. Following these criteria, we chose the ice-binding protein from the ocean pout, also called Type III Antifreeze protein (AFPIII), and the carbohydrate-binding module (CBM) of scaffolding CipA from *Clostridium Thermocellum*, CBM3a.

AFPIII is classified as an ice-binding protein (IBP) and an Antifreeze protein (AFP) synthesized in fish species exhibiting the unique ability to thrive in sea-ice environments without freezing. IBPs represent a growing category of proteins that assist organisms in surviving freezing conditions through various mechanisms involving adsorption to ice crystals. Within this group, AFP specifically refers to a subgroup of IBPs that can depress the freezing point of ice below its melting point, thereby impeding their growth and recrystallization processes.[21] AFP molecules display an irreversible adherence to ice crystals,[31,32] effectively impeding their growth and melting dynamics.[33–35]. Consequently, the presence of AFPs maintains the ice crystals at a reduced size and stable state for prolonged durations.[34] This binding mode may improve ice composites' ability to withstand temperature fluctuations and enhance their stability over time.

The particular AFPIII used in this study is a seven kDa protein free of cysteines and any post-translational modifications, with characterized ice-binding site and dynamics. Its structure is highly stable over time and can be produced recombinantly in *E. coli* in its active form, free or fused to other proteins such as GFP.[32] This protein has been one of our leading models in the past two decades in various biophysical studies.[32,36,37]

For the cellulose binding subunit of our chimera, we used the carbohydrate-binding module of scaffolding CipA from *Clostridium Thermocellum*, which belongs to family 3 CBM (CBM3a). The natural role of this protein family is to mediate the binding of the entire cellulosome to cellulose [22,38,39]. The structure of this protein is a beta-sandwich composed of two antiparallel beta sheets. A planar "strip" of aromatic amino acids exposed on one side of one of the beta sheets forms the cellulose-binding plane. These residues anchor the CBM3a to a single chain of crystalline cellulose through a series of hydrophobic interactions. Two additional polar amino acids bind glucose units from adjacent chains.[40]

We designed the chimera protein with the CBM3a subunit at the amino terminus of the protein and the AFPIII subunit at the carboxy terminus, separated by a linker (Figure 1a). A 6xHistidine tag after the AFPIII was used for protein purification. We cloned two chimera proteins that differed only in the linker region. In the first construct, we connected the CBM3a to the AFPIII with a minimal linker of only two amino acids, Glycine and Serine. In contrast, the second construct contained a ten amino-acid linker with the Glycine-rich sequence: 'GSGGGKGGGS.' This relatively long sequence adds both space and flexibility between the CBM3a and the AFPIII subunits of the chimera, adding translational freedom and potentially facilitating the interaction of each subunit with its relevant substrate. However, the longer linker may be more susceptible to protease degradation than the shorter one.

A general scheme explaining the production of the chimera protein is shown in Figure 1b. Both proteins were successfully expressed in *E. coli*. An SDS-PAGE analysis exhibited a strong band at approximately 30 kDa, consistent with the expected molecular weight of the chimera protein (Figure 1c). Activity analyses indicate that both chimera proteins are active and stable. Therefore, we continued the work with the potentially more stable short linker construct.

### 3.2     Large-scale production of CBM3a-AFPIII

The production of bioPykrete samples for mechanical testing requires hundreds of milligrams of CBM3a-AFPIII protein. To efficiently produce such quantities, we developed a fed-batch fermentation procedure. We obtained 40 g of cell dry weight per liter of fermented medium (CDW/L) and a final yield of 1 g of purified protein per 50 g of pellet, equivalent to 0.8 g of purified protein per 1 L of medium. This amount of protein is sufficient to obtain 0.5 L of CNC suspension containing 50 µM protein.

Engineered chimeras with dual functions, especially when originating from higher organisms such as fish, may be challenging to produce in bacteria due to multiple aspects such as folding, toxicity, and degradation. Nevertheless, AFPIII and CBM3a were previously expressed in *E. coli* in their soluble and active forms. Our chimera production was successful, and we obtained soluble protein without additional laborious procedures like refolding from inclusion bodies.

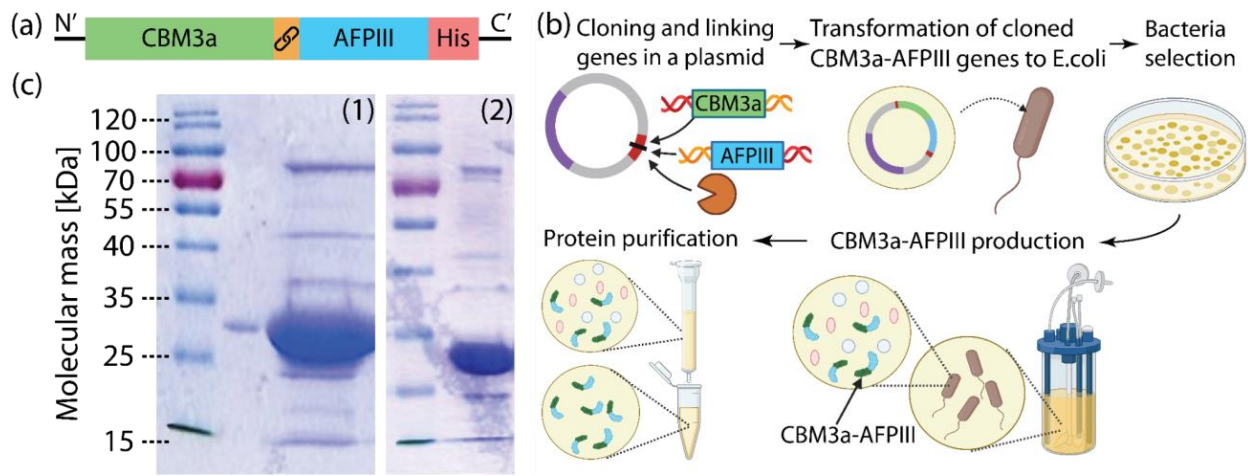

**Figure 1.** Production of the CBM3a-AFPIII chimera protein. (a) A schematic linear presentation of the protein subunits. The CBM3a gene is at the protein's amino terminus (N'), followed by a short or a long linker. The AFPIII gene follows, and the protein is terminated with a 6xHistidine-tag. (b) SDS-PAGE analysis of the chimera produced in flasks (panel 1) or by fermentation (panel 2), both after Ni-NTA affinity purification. (c) A schematic illustration of the production of the recombinant chimera protein. In practice, we inserted the two genes sequentially by cloning the AFPIII to a plasmid containing CBM3a.

### 3.3 Antifreeze activity studies

To verify that the AFPIII subunit in the CBM3a-AFPIII chimera was folded and active, we performed a set of assays commonly used to characterize the activity of AFPs. These tests include ice affinity purification, ice shaping, and thermal hysteresis assays. The results obtained with the longer linker construct are equivalent to those of CBM3a-AFPIII with the short linker and are not presented.

### 3.3.1 Ice affinity purification

Ice affinity purification methods are used to isolate and extract the active forms of IBPs [28,41,42] and to assess their ice-binding activity in a solution. We used the cold finger method in which ice is gradually grown (at a rate of up to 5 mm h$^{-1}$) on a metal finger immersed in the protein solution. Efficient mixing of the solution ensures that nonspecifically adsorbed impurities to the ice front are washed off and excluded, while the specifically adhering AFPIII molecules become engulfed by the ice. Figure S1 presents the SDS-PAGE analysis of the obtained ice, showing that residual non-AFP proteins, which were non-specifically purified by the Ni-NTA resin (visible in the SDS-PAGE in Figure 1c), did not bind to the ice.

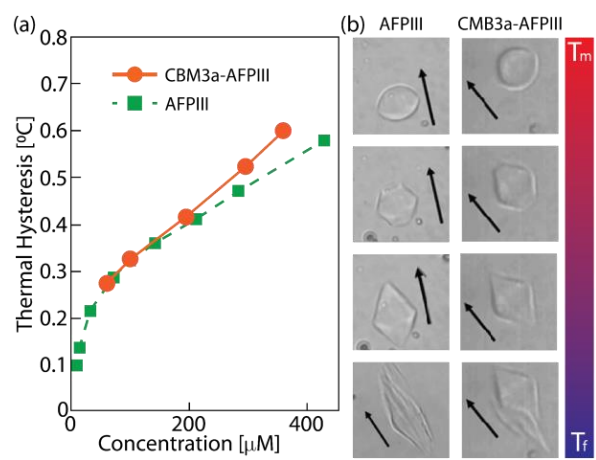

**Figure 2.** Ice-binding activities of CBM3a-AFPIII. (a) TH activity of CBM3a-AFPIII. Experimental results of the TH of the CBM3a-AFPIII chimera are compared to published data on AFPIII (adopted with permission from [43]), showing similar TH dependence on protein concentration. Data points represent the average of three independent experiments, with an error of less than 10% (b) Single ice crystal shapes stabilized by CBM3a-AFPIII, compared to ice shapes obtained in the presence of AFPIII (adopted with permission from [44]. The temperature scale bar shows the relative temperature regime in which the images were taken. The top row shows shaping during melting, *i.e.,* at temperatures slightly above the melting point. The bottom row shows the "burst" shape obtained just below the freezing point. The middle images show shaping between these two temperatures, *i.e.,* within the TH gap. The arrows denote the c-axis of the hexagonal ice crystals, which is also the growth/melt direction of ice crystals in solutions of AFP III.

### 3.3.2 Thermal Hysteresis

Thermal hysteresis (TH) of ice-containing IBPs is a phenomenon that arises from the adherence of these proteins to the faces of existing ice crystals and arresting their growth. [21] The result is a gap between ice crystals' melting and freezing points called the TH gap. While the effect of IBPs on the melting temperature is minor,[33,35] there is a pronounced inhibition of crystal growth at temperatures lower than the melting point. The TH gap is a strong indicator of ice-binding activity, and it can be quantified by measuring the melting and freezing points of a single ice crystal under a microscope. TH is highly dependent on protein type and concentration, with several orders of magnitude differences between various IBPs. We measured the TH of CBM3a-AFPIII and compared it to the TH gap of free AFPIII. Figure 2a shows that the TH of the chimera protein was indistinguishable at concentrations below 200 μM and slightly higher than the TH of AFPIII at higher concentrations, indicating that it retained, if not improved, the AFPIII ice binding affinity. The slight increase in the TH gap of the chimera can be explained by the Gibbs-Thomson effect, for which the larger molecules of the chimera protein (30 kDa) relative to the free AFPIII (8 kDa) can better resist engulfment, leading to an increase in TH. [45,46]

### 3.3.3 Ice shaping

Ice crystals in aqueous solutions at temperatures close to their freezing point grow and melt in a disk-like shape. However, in the presence of IBPs, the ice crystals adopt faceted shapes due to the binding of the proteins onto specific crystallographic planes. The specific spatial organization and chemical composition of the ice-binding sites of the proteins determine which planes are affected, as well as the mechanism and the dynamics of binding. Therefore, each IBP type produces unique crystal growth and melting shapes [44].

The ocean pout AFPIII produces bipyramidal crystals when the temperature of the crystals is lowered slightly below the melting point but remains within the TH gap. Initially, at temperatures slightly above the melting point, the tips of the bipyramids are melted, and all faces are rounded. When the temperature is lowered within the TH gap, the tips grow sharp along the c-axis of the crystal until the faces perpendicular to the c-axis vanish. At temperatures below the freezing point, a sudden growth of a large amount of ice spicules from the edges of the ice crystal occurs, retaining their orientation along the c-axis [44]. Figure 2b shows side-by-side the ice crystal shaping in the presence of either the CBM3a-AFPIII chimera or of free AFPIII. The visual similarity between the two ice crystals indicates that the chimera proteins, as desired, occupy the same planes as the free AFPIII and retain the same binding characteristics.

### 3.3.4 Cellulose binding

We analyzed the affinity of the chimera protein to CNC in a pull-down assay based on binding the protein to insoluble cellulose particles in a suspension and separating the soluble and insoluble fractions by centrifugation. We unfolded the protein with a denaturing agent to separate it from the cellulose and analyzed the bound and unbound fractions by SDS-PAGE. For control, we used AFPIII unlinked to CBM3a. As shown in Figure 3, the bound-to-CNC fraction of the CBM3a-AFPIII contained most of the protein used for the assay, while the unbound fraction was completely clear of protein. In the control assay, most of the protein remained in the unbound fraction. The small amount of protein noted in the bound fraction can be the protein that adsorbed non-specifically to the cellulose and was not washed out efficiently or a protein leak during gel loading due to overload. These findings imply that the chimeric protein binds efficiently to CNC, while protein without the CBM3a subunit does not. The presence of only a single band suggests that the two subunits of the chimera remain intact.

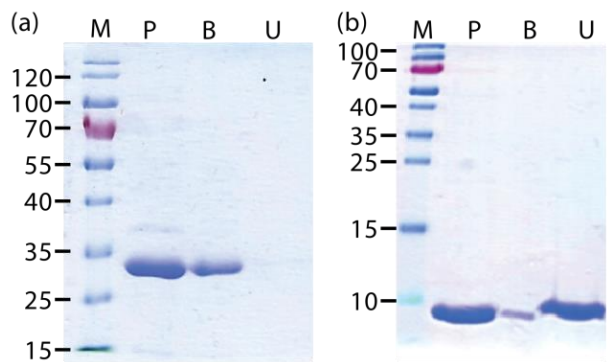

**Figure 3.** Cellulose binding affinity of CBM3a-AFPIII. M- Marker. P- Purified protein before assay. B- Cellulose bound fraction. U- Unbound fraction. (a) CBM3a-AFPIII (b) AFPIII only.

### 3.4    BioPykrete fabrication

We developed a CNC-ice composite fabrication procedure that ensures directional and uniform ice growth, which is essential for the uniformity of the compression models. This work focused on the compression strength perpendicular to the ice growth direction. This orientation is expected to be the strongest and, therefore, the most relevant for load-bearing applications.

Figure 4a illustrates a cooling setup scheme with a temperature-controlled cooling stage on the bottom and a sealed chamber on top. Different CNC suspensions were poured into specially designed cups and placed inside the chamber, allowing for the simultaneous production of five samples for mechanical compression testing. Each cup featured a copper bottom for efficient heat conduction from the cold base of the cooling stage, while its insulated body ensures unidirectional heat flux and promotes unidirectional growth of ice. We cool the suspension from the bottom at a constant heat flux by reducing the base temperature at a constant rate, compensating for the increase in the distance of the ice front from the cold base[30]. Figure S2 shows the directional freezing setup and typical composite samples.

As the ice front advances, the newly formed ice crystals exclude the suspended CNC particles. Figure 4b-4c presents how they (the CNC particles) link to create a network of elongated cells that trap the ice inside. Figure 4d illustrates how the chimera proteins link the CNC to ice and help regulate its growth.

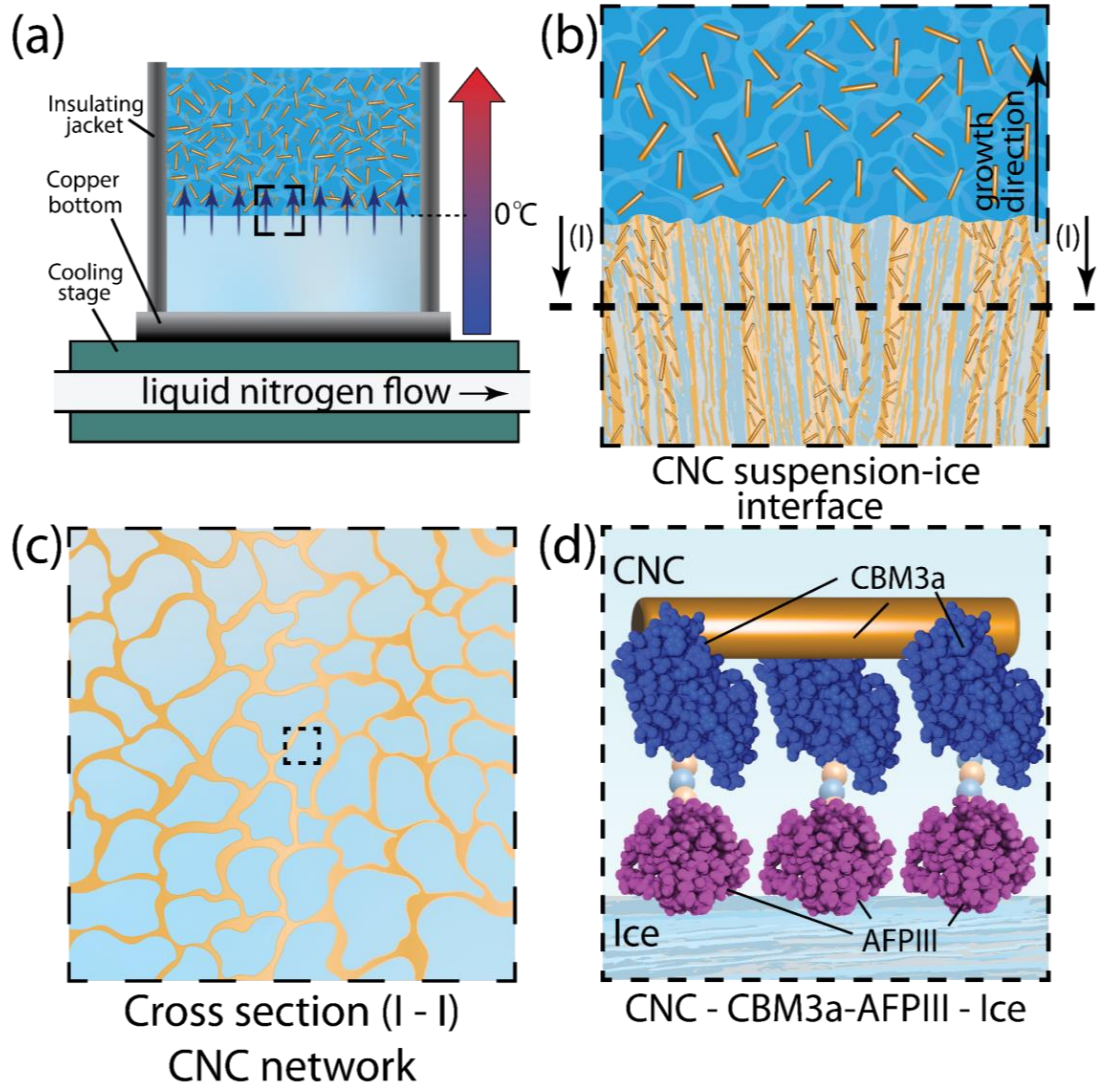

**Figure 4.** BioPykrete. (a) An illustration of the directional cooling setup. The cooling rate of the solution is controlled by the flow rate of liquid nitrogen through the bottom plate. (b) An illustration of the ice-CNC composite formation. The CNC suspension is cooled from the bottom and insulated from the sides to prompt unidirectional (1D) ice growth. The advancing solidification front expels the CNC particles, forming a network of elongated cells that encapsulate the ice. (c) An illustration of cross-section (I-I) in 4(b) of the CNC network. (d) An illustration of the chimera protein aiding to bind the CNC network to the ice and regulate its growth.

## 3.5 Elastic energy and Morphological Properties of CNC-ice composites

In the brittle ice, fractures can propagate unimpeded, resulting in an abrupt failure at low stress. On the other hand, in traditional brittle composite materials (BMCs), additive fibers are used to improve the mechanical properties of the bulk. With the fibers entangled in the brittle matrix, propagating cracks are forced to change directions constantly and overcome the additional energy required to pull the fibers apart. As a result, the global failure mode changes from abrupt to gradual [8], and the mechanical response of the composite material appears almost "elasto-plastic" in nature. In our case, these are not added fibers but rather CNC particles that self-organize into a network of sub-millimeter cells that encapsulate the ice within (Figure 5, inset). Although cracks can still form in the ice, they are now confined to typical pore size, and additional work is required to displace, deform, and tear the CNC network relative to the ice. Therefore, the quality of the CNC network becomes the determining factor for the composite's strength. Although cracks might also propagate in the CNC network, this is highly unlikely due to the significant mismatch in the elastic modulus and the stress-to-failure between pure ice and CNC structures [47].

We used mechanical compressive strength testing to quantify the global effects of the additive proteins on the resulting stress and energy to failure of the macrostructure. Figure 5 presents a 70-fold improvement in volume-average specific energy-to-failure from pure ice (defined at a strain of 0.07) to BioPykrete (defined at a strain of 0.3).

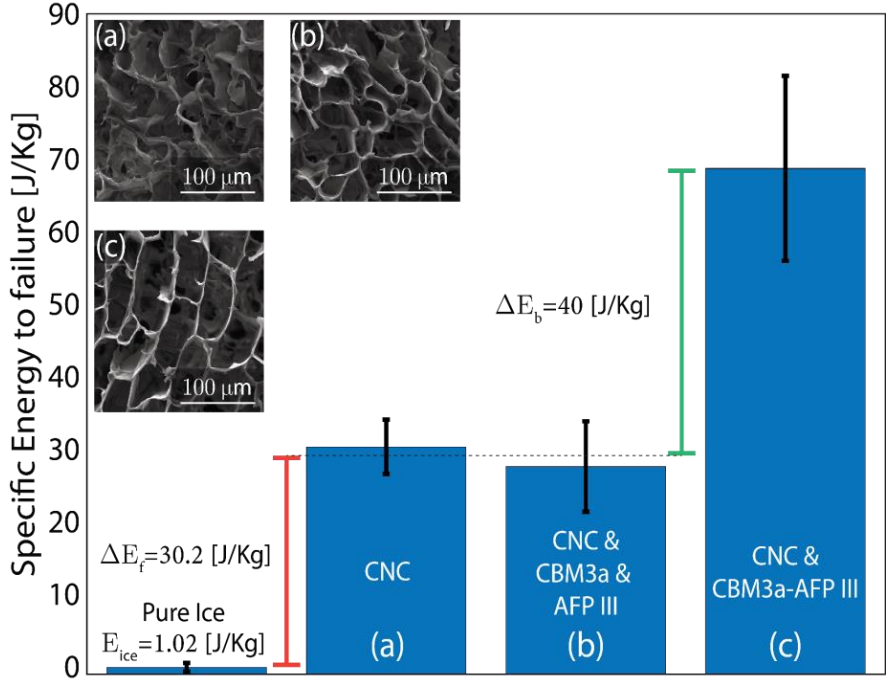

**Figure 5.** Comparison between the average energy-to-failure of the different ice composites, analyzed from the compression tests. The energy to "failure" was taken as the energy required to reach a strain of f $\varepsilon = 0.3$ for all the samples except for the pure ice, which failed at a significantly lower strain ($\varepsilon = 0.07$). The error bars are 95% confidence over the mean. Insets (a-c) present SEM images of the typical pore structure for the three composite materials.

We can attribute this increase to the additional network strength (N) and bonding energy (B).

$$E_{tot} = E_N + E_B$$

Assuming only the chimera protein in the BioPykrete contributes to the bonding energy, we find $E_N = 30.2 \, J/Kg$ and that $E_B = 40 \, J/Kg$. It is worth noting that adding the chimera protein more than doubles the energy-to-failure of the entire sample. This is a synergistic effect where the chimera protein augments the entirety of the CNC network, leading to an overall global improvement of the mechanical properties.

The inset in Figure 5 shows the cross-sections of the three new types of bio-ice composite materials. Two trends can be observed, the first is a morphological change with the addition of unbonded CBM3a and AFPIII to the CNC–ice suspension. This results in the most drastic effect: the pores become regular with stable "walls". A possible explanation for the change in material morphology is that the fast-binding kinetics of AFPIII to ice prism planes [37] influence the ice growth such that fingering patterns become more pronounced, producing sharper and more uniformly shaped crystals. This is consistent with the sharp elongated crystal shapes obtained in AFPIII solutions [44]. As presented in Figure 5b-5c, this change in the microstructure has little to no effect on the energy to failure in our directional compression tests. However, the finer and better-pronounced microstructure limits the size a propagating crack can reach before encountering a CNC barrier that hinders its advance. As a result, the stress-strain curves are smoother in Figure 6c compared to 6b.

The second trend occurred when the chimera protein CBM3a-AFPIII was added to the CNC suspension. The pores become more uniform in size and shape i.e. regular. The CNC "walls" appear to have fewer defects and smudges from cutting the samples in preparation for SEM imaging. This implies that the self-organized CNC network has become more mechanically robust. Indeed, even though the pore morphology changed only slightly, the energy to failure increased drastically.

### 3.6 Mechanical properties of CNC-ice composites

Figure 6a shows the behavior of pure ice, which is our point of reference. The ice exhibits abrupt failure at relatively low stress with a typical saw-tooth stress-strain curve. This is likely the result of several large fractures propagating through most of the matrix. The unpredicted behavior of these cracks is the underlying reason for the dispersion in maximum strength and strain to failure of pure ice, as each one of the propagating fractures can result in a critical failure of the bulk. The addition of CNC to the ice increases the stress to failure fivefold and changes the mode of failure from a brittle - sudden and abrupt - to a smoother "elasto-plastic" like response to stress. As apparent in Figure 6b, we can still identify the onset of large fractures, appearing as "kinks" in some of the measured stress-strain curves. However, in a typical BMC manner [8], they stop instead of resulting in a catastrophic failure of the specimen. We note that the CNC suspension used contained 0.3 mM Tris-HCl due to buffering of our original CNC suspension. Tris-HCl is not an ice-active substance, and it is reasonable to assume its effects on ice properties are negligible at such low concentrations. In addition, ice grown from fresh water is considerably stronger than ice grown from salt water [48], suggesting that any effect caused by the buffer would weaken the ice, further strengthening our results. With the addition of unbounded AFPIII and CBM3a (Figure 6c), the network becomes more regular, although not stronger. The resulting stress-strain curves are more uniform, with less apparent kinks and show a distinct initial elastic-like phase and a final plastic-like phase. According to BMCs, the initial phase is explained by forming an effective global energetic barrier on the microstructure level that cracks must overcome to propagate in bulk. On a larger scale, the propagating crack has to break the CNC network from the ice to advance. Such global processes are much slower than the propagation speed of a crack in pure ice. Indeed, instead of a crash in the form of a sawtooth in the later phase, we observe a more gradual and slower response to (above critical) stress. Finally, introducing the chimera protein as a bio-adhesive capable of bonding the CNC to ice crystals improves the overall strength of the composite by an additional factor of 2, to a total 10-fold increase relative to pure ice. Moreover, as shown in Figure 6d, the additional bonding energy barrier leads to smoother stress-strain curves.

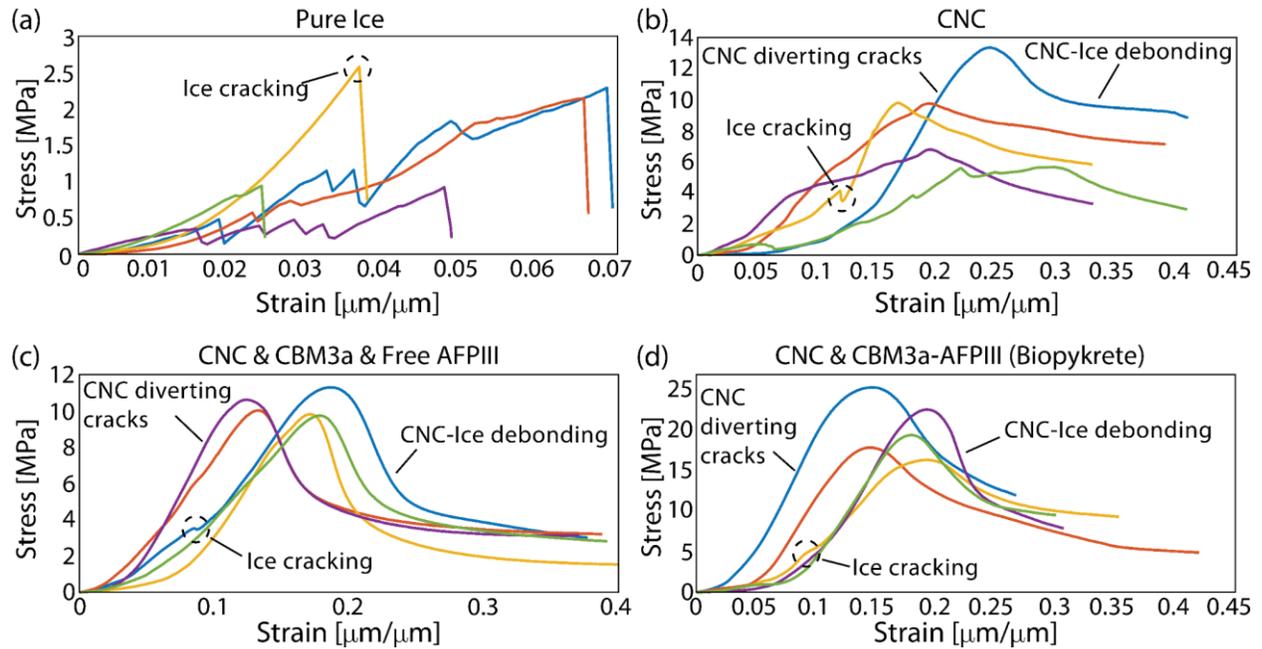

**Figure 6.** Experimental results of compression tests performed on CNC-ice composites. For each composite type, 5-6 independent experiments were conducted, and all repeats are presented. (a) Pure ice exhibits a classical abrupt brittle failure due to fracture propagation. (b) The added CNC self-organizes into a network of cells, encapsulating the ice and changing the composite's failure mechanism. However, this microstructure is too weak to suppress all propagating fractures, as evident from the "kinks" in the stress-strain curves, some marked by dashed circles to guide the eye. (c) The addition of unbound AFPIII and CBM3a acts to smooth out the curves and reduce the number of "kinks", but has little effect on the strength or elongation of the new composite. (d) BioPykrete. The additional bonding energy due to the chimera protein bio-adhesive properties increases the sample strength by an additional 2-fold to a total increase of 10-fold in compressive strength compared to pure ice.

## 4. Conclusions

Herein, we present the development of BioPykrete, an ice composite with superior mechanical properties. To create the BioPykrete, we reinforced ice with a self-organizing network of nanocrystalline cellulose (CNC) bound to it by a chimera protein acting as the bioadhesive. The chimera protein is a tailor-made fusion between the ice-binding AFPIII and the cellulose-binding CBM3a proteins. We detail the chimera development process, including the move to large-scale production and results validation.

We produced several different composites, inspected their microstructure, and tested their compressional strength and energy to failure. BioPykrete showed a 10-fold increase in compressional strength and has more than 70 times the energy-to-failure compared to standard ice. Moreover, the failure mechanism changed from a sudden and brittle snap to a ductile behavior, a crucial improvement for engineering applications. We hypothesize that this is due to CNC self-organizing into a 3D network of cells enclosing the ice within, augmented, and regulated by the chimera protein. In this network, cracks cannot propagate freely, i.e. grow to a size above a typical pore size before overcoming a significant energetic barrier. The resulting bio-composite material has a compression strength comparable to standard concrete [49] and a failure mechanism favorable for engineering applications, such as construction.

A recent study of ice reinforcement by PVA and a series of fibers show another example of significant ice reinforcement by an ice-active substance (PVA) in addition to fibers. [50] Yet, our study offers a sustainable solution that utilizes a unique molecule tailor-made for ice reinforcement. A combination of our chimera molecule with other ice-active substances would likely reinforce the ice composites. Noteworthy, combining other chimeras using different types of antifreeze and cellulose-binding proteins (AFPs and CBMs) may offer even stronger reinforcement. Such a system requires significant optimization.

Biopykrete has a mechanical response similar to brittle composite materials (BCMs), a general model that describes the effects of embedded fibers on the strength of brittle materials. Therefore, a more advanced strength and fracture dynamics analysis, capable of considering the internal structure of the CNC network, is needed to help guide future improvements and optimization. Combined with tools like high-speed x-ray tomography that can image the complex 3D motion of cracks in real-time, such research can offer new insights to the field of fracture mechanics in bio-composites. Excitingly, as BioPykrete possesses mechanical properties comparable to concrete, it can be tested in real-life construction applications in the Arctic or Antarctic circles.

Our results can pave the way for a wide range of future research opportunities. For example, a deeper investigation into the solidification kinetics could explain the interplay between the CNC, the water, and the chimera protein. Such understanding could drive an optimization process of the antifreeze and cellulose-binding proteins used to construct the chimera, in the solution composition, as well as in the freezing protocol itself. Furthermore, directly incorporating biological building blocks into existing technologies is a promising path for the future of composite materials. Protein engineering can be used to generate chimeras with tailored binding affinities, allowing not only to strengthen composite materials mechanically but also to adjust and regulate their properties, as well as to introduce new features. Elements like the chimera protein created in this work allow for complex manipulations on the angstrom to nano scales that are otherwise very hard to achieve.

**Supporting information is available.**

**Acknowledgments**

We thank the Israel Science Foundation for the financial support (I.B for Grant No. 1308/21 and V.Y for Grants No. 838/23 and 2044/23). We thank Dr. Einat Zelinger (Agricultural Core Facility, HUJI) for assistance with SEM and Cryo-SEM.

# Biomimetic Engineering of a Fortified Ice Composite with Enhanced Mechanical Properties

## Supplementary Information

Chen ADAR[1#], Yulia BARON[1#], Baruch ROFMAN[3#], Maya BAR-DOLEV[1,2], Liat BAHARI[1], Victor YASHUNSKY[4,1], Vera SIROTINSKAYA[1], Oded SHOSEYOV[5], Ido BRASLAVSKY[1]*

1 Institute of Biochemistry, Food Science, and Nutrition, Robert H. Smith Faculty of Agriculture, Food and Environment, The Hebrew University of Jerusalem, Rehovot 7610001, Israel

2 Faculty of Biotechnology and Food Engineering, Technion, Haifa 3200003, Israel

3 Swiss Plasma Center, Ecole Polytechnique Fédérale de Lausanne, Lausanne CH-1015, Switzerland

4 The Swiss Institute for Dryland Environmental and Energy Research, The Jacob Blaustein Institutes for Desert Research, Ben-Gurion University of the Negev, Sede Boqer Campus, 84990 Midreshet Ben-Gurion, Israel

5 Department of Plant Sciences and Genetics in Agriculture, Robert H. Smith Faculty of Agriculture, Food and Environment, The Hebrew University of Jerusalem, Rehovot 7610001, Israel

## Methods

1. **Engineering the CBM3a-AFPIII chimeras**

The CBM3a-AFPIII fusions were constructed by cloning the AFPIII gene downstream to the gene coding for CBM3a in the pET-28a plasmid. Our starting point was a plasmid bearing the CBM3a sequence and a 6 x Histidine tag at the C'. This plasmid was previously cloned by the group of Ed Byer (Weizmann Institute of Science, Rehovot, Israel) and used to express the CBM3a.

The gene coding for AFPIII was amplified by Polymerase Chain Reaction (PCR) using a plasmid bearing the gene obtained from P.L. Davies (Queen's University, Kingston, Canada) as a template. The Forward primers used for the PCR reaction were 5'-AAGGATCCAACCAGGCTAGCGTTGTGGCC-3' for the short, two amino-acid linker and

5'-AAGGATCCGGCGGCGGCAAAGGCGGCGGCTCTAACCAGGCTAGCGTTGTGG-3',

for the long, glycine-rich 8-amino acid linker (LL8). The reverse primer for both reactions was 5'-TTCTCGAGAGCAGCGTAACCTTTAACCATG-3'.

PCR was performed using OneTaq DNA polymerase (New England BioLabs, MA, USA) and the buffer supplemented by the enzyme manufacturer. The reaction was as follows: preheating at 94 °C for 1 min, four cycles at 94 °C for 30 sec, 48 °C for 30 sec and 68 °C for 30 sec, 26 cycles at 94 °C for 30 sec, 60 °C for 30 sec and 68 °C for 30 sec, followed by final elongation period at 68 °C for 5 min. The amplified products were recovered from 0.8% agarose gel using the PCR purification kit (Qiagen, Germany). The PCR products and the pET28a plasmid were digested with restriction endonucleases BamHI and XhoI and ligated into pET-28a using T4 DNA ligase (New England BioLabs, MA, USA).

## CBM3a-AFPIII (short linker version) gene:

ATGGCAAATACACCGGTATCAGGCAATTTGAAGGTTGAATTCTACAACAGCAATCCTTCAGATACTACTA
ACTCAATCAATCCTCAGTTCAAGGTTACTAATACCGGAAGCAGTGCAATTGATTTGTCCAAACTCACATT
GAGATATTATTATACAGTAGACGGACAGAAAGATCAGACCTTCTGGTGTGACCATGCTGCAATAATCGG
CAGTAACGGCAGCTACAACGGAATTACTTCAAATGTAAAAGGAACATTTGTAAAAATGAGTTCCTCAAC
AAATAACGCAGACACCTACCTTGAAATAAGCTTTACAGGCGGAACTCTTGAACCGGGTGCACATGTTCA
GATACAAGGTAGATTTGCAAAGAATGACTGGAGTAACTATACACAGTCAAATGACTACTCATTCAAGTCT
GCTTCACAGTTTGTTGAATGGGATCAGGTAACAGCATACTTGAACGGTGTTCTTGTATGGGGTAAAGAA
CCCGGTGGCAGTGTAGTACCATCAACACAGCCTGTAACAACACCACCTGCAACAACAAAACCACCTGCA
ACAACAATACCGCCGTCAGATGATCCGAATGCAGGATCCAACCAGGCTAGCGTTGTGGCCAACCAGCTG

ATCCCGATTAATACTGCTCTGACTCTGGTTATGATGCGTAGTGAAGTTGTTACTCCGGTTGGTATCCCGG
CCGAAGATATCCCGCGTCTGGTTAGCATGCAGGTTAACCGTGCTGTTCCGCTGGGTACCACTCTGATGCC
GGACATGGTTAAAGGTTACGCTGCTCTCGAGCACCACCACCACCACCACTGA

### CBM3a-AFPIII (short linker version) protein sequence:

MANTPVSGNLKVEFYNSNPSDTTNSINPQFKVTNTGSSAIDLSKLTLRYYYTVDGQKDQTFWCDHAAIIGSN
GSYNGITSNVKGTFVKMSSSTNNADTYLEISFTGGTLEPGAHVQIQGRFAKNDWSNYTQSNDYSFKSASQFV
EWDQVTAYLNGVLVWGKEPGGSVVPSTQPVTTPPATTKPPATTIPPSDDPNAGSNQASVVANQLIPINTAL
TLVMMRSEVVTPVGIPAEDIPRLVSMQVNRAVPLGTTLMPDMVKGYAALEHHHHHH

### CBM3a-LL8-AFPIII (long linker version) gene:

ATGGCAAATACACCGGTATCAGGCAATTTGAAGGTTGAATTCTACAACAGCAATCCTTCAGATACTACTA
ACTCAATCAATCCTCAGTTCAAGGTTACTAATACCGGAAGCAGTGCAATTGATTTGTCCAAACTCACATT
GAGATATTATTATACAGTAGACGGACAGAAAGATCAGACCTTCTGGTGTGACCATGCTGCAATAATCGG
CAGTAACGGCAGCTACAACGGAATTACTTCAAATGTAAAAGGAACATTTGTAAAAATGAGTTCCTCAAC
AAATAACGCAGACACCTACCTTGAAATAAGCTTTACAGGCGGAACTCTTGAACCGGGTGCACATGTTCA
GATACAAGGTAGATTTGCAAAGAATGACTGGAGTAACTATACACAGTCAAATGACTACTCATTCAAGTCT
GCTTCACAGTTTGTTGAATGGGATCAGGTAACAGCATACTTGAACGGTGTTCTTGTATGGGGTAAAGAA
CCCGGTGGCAGTGTAGTACCATCAACACAGCCTGTAACAACACCACCTGCAACAACAAAACCACCTGCA
ACAACAATACCGCCGTCAGATGATCCGAATGCAGGATCCGGCGGCGGCACCGGCGGCGGCTCTAACCA
GGCTAGCGTTGTGGCCAACCAGCTGATCCCGATTAATACTGCTCTGACTCTGGTTATGATGCGTAGTGAA
GTTGTTACTCCGGTTGGTATCCCGGCCGAAGATATCCCGCGTCTGGTTAGCATGCAGGTTAACCGTGCTG
TTCCGCTGGGTACCACTCTGATGCCGGACATGGTTAAAGGTTACGCTGCTCTCGAGCACCACCACCACCA
CCACTGA

## CBM3a-LL8-AFPIII (long linker version) protein sequence:

MANTPVSGNLKVEFYNSNPSDTTNSINPQFKVTNTGSSAIDLSKLTLRYYYTVDGQKDQTFWCDHAAIIGSN
GSYNGITSNVKGTFVKMSSSTNNADTYLEISFTGGTLEPGAHVQIQGRFAKNDWSNYTQSNDYSFKSASQFV
EWDQVTAYLNGVLVWGKEPGGSVVPSTQPVTTPPATTKPPATTIPPSDDPNAGSGGGTGGGSNQASVVA
NQLIPINTALTLVMMRSEVVTPVGIPAEDIPRLVSMQVNRAVPLGTTLMPDMVKGYAALEHHHHHH

2. **CBM3a-AFPIII production and purification**

    **2.1 Production in flasks**

The recombinant proteins with the short and the long linkers were expressed in *Escherichia coli* BL21 cells in 2 L flasks containing 0.5 L Luria-Bertani (LB) broth, supplemented with 100 μg L$^{-1}$ of kanamycin at 37 °C to mid-exponential phase (OD$_{595}$~0.6). The bacterial culture was induced by adding one mM isopropyl-β-D-thiogalactopyranoside (IPTG) and cultivated at 16 °C overnight. Recombinant *E. Coli* cells were harvested by centrifugation at 6000 rpm and four °C for 15 min before being resuspended in 40 mL Binding Buffer (5 mM imidazole, 50 mM Tris-HCl, 150 mM NaCl, pH 7.6). The suspensions were kept at -80 °C until use. To extract the protein from the bacteria, we added 1 mL of a homemade protease inhibitor cocktail containing 23 mM AEBSF (inhibits serine proteases), two mM Bestatin (inhibits aminopeptidases), 0.3 mM Pepstatin (inhibits acid proteases), and 0.3 mM Leupeptin (inhibits cysteine proteases). The suspension was cooled by placing the tube in an ice bucket and sonicated in 10 pulses of 30 s on/30 s off (Sonics Vibra-cell Model CV33). The cell extract was maintained at four °C while centrifuged at 13000 rpm for 30 min before the lysate was collected. A protease inhibitor cocktail was added again.

    **2.2 Large-scale protein production (fermentation)**

Recombinant CBM3a-AFPIII protein was expressed in a stirred tank fermenter of 1 L or 5 L (Applikon Biotechnology, The Netherlands) for a fed-batch fermentation process. The fermentation procedure's development was described in detail (Sirotinskaya et al. 2024). The bacteria were grown under controlled physical and chemical parameters (temperature, airflow, oxygen flow, agitation speed, foam formation, pH, and dissolved oxygen). The feed and medium components, e.g., carbon and nitrogen sources, salts, etc., were supplied by direct injection into the vessel during the fermentation process. Air was provided at a rate of 0.15-2 vvm (volume of air per volume of medium per minute) to maintain the dissolved oxygen concentration greater than 30% of air saturation. pH 7 was maintained by titration of either 2 M H2SO4 or 3 M NaOH. The cultivation temperature was 37 °C until the optical density at 595 nm ($OD_{595}$), indicative of cell concentration, reached 0.6. At this point, the temperature was lowered to 15-20 °C, and the culture was supplemented with one mM of IPTG to induce protein production in excess. The typical fermentation duration was 20-24 hours, and final $OD_{595}$ values reached 60 to 80. Cells were harvested by centrifugation at four °C at 4500 rpm for 45 min and stored at −80 °C until purification. Frozen harvest was resuspended in binding buffer (5 mM imidazole, 50 mM Tris-HCl, 150 mM NaCl, pH 7.6) at a ratio of 10 mL buffer per 1 g of pellet, precooled to 4 °C and supplemented with a protease inhibitor cocktail as described above.

### 2.3 Protein purification

Recombinant proteins were purified from the lysates using Ni-NTA affinity purification. 1 mL of Ni-NTA agarose beads suspension was added per 1 gr of pelleted culture and incubated for 1 hr at four °C. The solution was placed in a gravity column, and the lysate flowed out. The beads were washed with 5 column volumes (CV) of binding buffer and eluted in a two-step procedure: 5 CV of an elution buffer containing 50 mM Tris-HCl, 150 mM NaCl, and 100 mM Imidazole (pH 7.6), and then 5 CV of the same elution buffer but with 250 mM Imidazole. Fractions were collected and analyzed by 12% w/v SDS PAGE. Fractions containing purified protein (determined by the presence of a significant band at 29 kDa, corresponding to CBM3a-AFPIII) were pulled and dialyzed against 50 mM Tris, pH 7.6. Protein was concentrated using Vivaspin ultrafiltration columns (Sartorius), and the final protein concentration was determined using the micro-BCA Protein Assay (Thermo Scientific, cat#23235). Overall, 937 mg of CBM3a-AFPIII was purified from 50 gr of bacterial pellet. The control proteins, AFPIII and CBM3a, were prepared following the same procedure of expression and purification, with a modification of the cultivation, which was carried out at 37 °C for 4-6 hours.

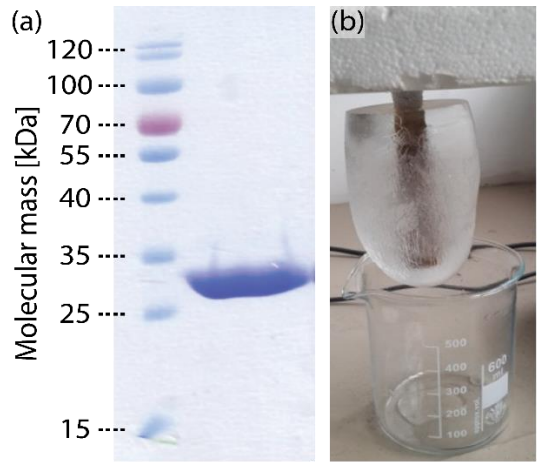

**Figure S1**. Ice affinity purification of CMB3a-AFPIII. (a) SDS-PAGE analysis of the ice fraction shows that the CMB-AFPIII successfully and selectively bonded to the ice. Before the ice affinity purification, the solution contained some other proteins, as shown in the main text in Figure 1c. (b) A picture of the purified ice collected on the cold finger at the end of the purification process.

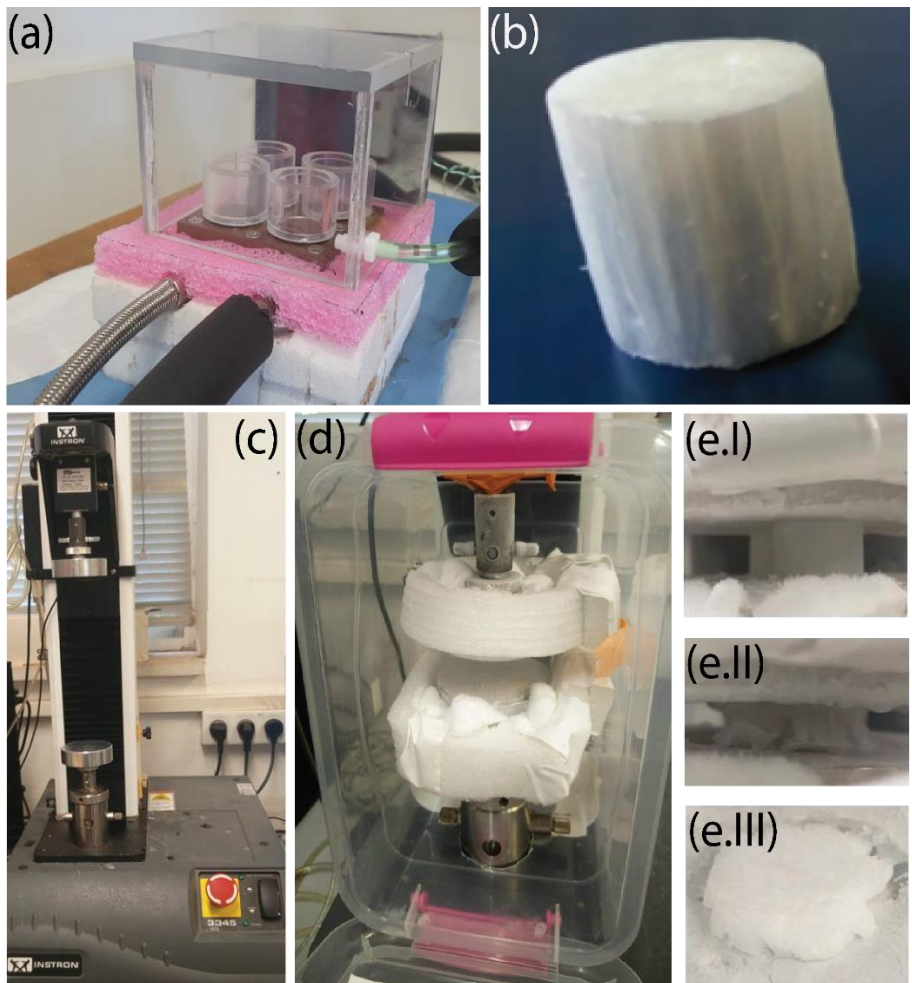

**Figure S2**. Experimental setups for sample preparation and compression tests. (a) The directional freezing insulated chamber can prepare four samples at once. A controlled flow of liquid nitrogen cools the bottom copper stage. (b) A typical sample with a diameter of 5 cm and a length of 10 cm, is ready for compression testing. Experimental setup for compressional testing. (c) The INSTRON that was used for compression tests. (d) the insulated chamber used for compression tests at low temperatures. Dry ice was supplied to keep the temperature low. (e) CNC-ice sample during a representative compressive test. The sample is aligned parallel to the ice growth direction. (e.I) At the beginning of the compression test. (e.II) During compression. (e.III) At the end of the compression test.